\documentclass[namedreferences]{SolarPhysics}
\usepackage[optionalrh]{spr-sola-addons} 
\usepackage{graphicx}                    
\usepackage{color}                       
\usepackage{url}                         

\newcommand{\aap}{    {\it Astron. Astrophys.}}

\newcommand{\apj}{    {\it Astrophys. J.}}

\newcommand{\grl}{    {\it Geophys. Res. Lett.}}

\newcommand{\jgr}{    {\it J. Geophys. Res.}}

\newcommand{\solphys}{{\it Solar Phys.}}

\newcommand{\ssr}{    {\it Space Sci. Rev.}}

\begin{document}

\begin{article}

\begin{opening}

\title{Estimated relations at a shock driven by a coronal mass ejection}

%
\author{M.~\surname{Eselevich}$^{1}$\sep
        V.~\surname{Eselevich}$^{1}$
       }

%
\runningauthor{Eselevich and Eselevich}

\runningtitle{Estimated relations at a shock}

%
  \institute{$^{1}$ Institute of solar-terrestrial physics,
  Irkutsk, Russia \\
  email: \url{mesel@iszf.irk.ru}
  email: \url{esel@iszf.irk.ru}
             }

\begin{abstract}
Analysis of SOHO/LASCO C3 data reveals a discontinuity,
interpreted as a shock wave, in plasma density radial profiles in
a restricted region ahead of each of ten selected coronal mass
ejections (CME) along their travel directions. In various events,
shock wave velocity $V\approx$ 800-2500 km s$^{-1}$. Comparing the
dependence of Alfv\'{e}n Mach number $M_A$ on shock wave strength
$\rho_2/\rho_1$, measured at $R > 10R_\odot$ from the center of
the Sun, to ideal MHD calculations suggests that the effective
adiabatic index $\gamma$, characterizing the processes inside the
shock front, is largely between 2 and 5/3. This corresponds to the
effective number of degrees of freedom of motion 2 to 3. A similar
dependence, $M_A(\rho_2/\rho_1)$, was derived for the Earth's bow
shock and interplanetary collisionless shock waves. All this
supports the assumption that the discontinuities in front of CMEs
are collisionless shock waves.
\end{abstract}

%

\end{opening}

%
\section{Introduction}

If the Rankine-Hugoniot relations are valid at the discontinuity
in supersonic plasma (or gas) stream parameters, the discontinuity
may be interpreted as a shock wave. Even to roughly estimate the
validity of these relations for a shock discontinuity driven in
front of fast CMEs is no easy task. This is first of all
associated with the fact that a shock front is difficult to
identify and register in the coronal conditions. Nevertheless,
several such attempts for individual events have been made in
\opencite{Vourlidas2003}; \opencite{Eselevich2007};
\opencite{Manchester2008}; \opencite{Eselevich2008};
\opencite{Ontiveros2009} (hereafter Paper 1).

Two different approaches may be singled out which have been
applied to identifying a shock front on coronal images. The first
one is numerical simulation within the framework of the ideal
magnetohydrodynamics (MHD) of the entire process of CME formation
and propagation. In this case, the manner in which the CME is
formed may be arbitrary. The results of such a simulation were
compared to the experiment in order to identify the shock front
\cite{Vourlidas2003,Manchester2008}. Note that in this case one
can directly estimate the relation between thermodynamic
parameters at the discontinuity because they are included in the
model.

The second approach relies on the fact that there is no shock
discontinuity if the velocity of the generating CME remains below
the critical velocity, $u_C$; while such a discontinuity does
exist at supercritical velocities. Regardless of the difficulty in
estimating the critical velocity in the corona, this approach
enabled us to demonstrate that the transition does exist, and
consequently the observed discontinuity is a shock wave
\cite{Eselevich2008}. There is some difficulty in identifying the
shock fronts in coronal images stemming from the fact that the CME
frontal structure and the shock front in front of it are identical
geometrically and are often rather close to each other. We
employed two different complementary methods to make sure that the
observed discontinuity in the brightness distribution is the shock
front rather than the CME frontal structure behind it:
\begin{enumerate}
\item We examined the conditions in the disturbed zone in front of CMEs
with different velocities $u$. For CMEs with velocities exceeding
$u_'$, the shock-wave-related discontinuity was observed in the
frontal part of a rather extended disturbed zone.
\item The processes of the disturbed zone evolution and shock wave
formation were investigated for some CMEs, in the coordinate
system associated with their frontal structures, when the CME
velocity $u$ passes the critical velocity $u_C$
\cite{Eselevich2010}.
\end{enumerate}
Note that the value of the critical velocity at which a shock wave
formed appeared to be rather close to existing estimates of the
Alfv{\'e}n velocity in the corona \cite{Mann1999}. These methods
allowed the shock front width to be 
measured for several tens of CMEs at various distances (up to
$6R_\odot$) from the center of the Sun ($R_\odot$ is the Sun's
radius).

Thus, whether by one or another method, we can identify a shock
wave in the corona; so the next obvious step would be to try to
verify the validity of the Rankine-Hugoniot relations at the
shock. We are restricted, however, in that we cannot measure just
{\it any} plasma parameter in the corona. White-light coronal
images allow the shock discontinuity velocity to be determined
(from the change in the discontinuity position with time) and the
density ratio estimated at the discontinuity, relying on some
assumptions. Apparently, it is these values that should be taken
as a basis for the analysis. Moreover, similarly to the Earth's
bow shock, we may apply simple relations of the ideal MHD at a
flat shock discontinuity to investigate shocks in the corona, i.e.
make use of the assumption that the shock front width is smaller
than the radius of its curvature.

The aim of this work is to obtain Rankine-Hugoniot relations at
the flat shock discontinuity, within the framework of the ideal
MHD, and to estimate their validity for a number of identified
shocks driven ahead of CMEs.

\section{Data and method of analysis}

The analysis involved LASCO C3 coronal images with a 4 to 30
$R_\odot$ field of view \cite{Brueckner1995}. The images contain
many stars that are much brighter than the corona. Such point
objects as stars can be easily removed by a sigma filter. The
filter is applied to calculating the mean value and rms deviation
in a square neighborhood centered at each image pixel, excluding
the pixel itself. If the brightness of the central pixel exceeds
the mean value by more than a certain value, it is replaced by the
mean. The filtering of all images involved a 15$\times$15 pixel
neighborhood and a maximum deviation equaling three rms
deviations, while the filtering procedure was iterated ten times.
After filtration, the images of bright stars are effectively
eliminated, the coronal brightness signal remaining almost
unchanged.

The images thus processed were presented as difference brightness
$\Delta P = P(t) - P(t_0)$, where $P(t)$ is the coronal brightness
at instant $t$, corresponding to the event in question, $P(t_0)$
is the undisturbed brightness at fixed instant $t_0$ chosen well
before the event. Thus, the background brightness due to the
stationary distribution of plasma in the corona was eliminated.
Calibrated LASCO images with full brightness $P(t)$ (in units of
mean solar brightness, $P_{MSB}$) enable us to use the difference
brightness to estimate the mean density of the matter causing the
change in brightness, if we make an assumption with regard to its
spatial size along the line of sight.

The difference brightness was used to examine the dynamics of
CMEs, as well as of the disturbed zone and shock wave in front of
them. It has proved to be more convenient to employ maps of
difference brightness isolines rather than the traditional
grey-scale images for selecting and generally analyzing events.
The arrangement of isolines provides for a better understanding of
the character of the difference brightness distribution, direction
and value of the gradients. Isolines levels were selected
depending on the event and the distance at which it was
considered.

It is more convenient to make quantitative measurements using
difference brightness profiles drawn in a certain direction. For
this purpose, we used radial profiles $\Delta P(R)$ of difference
brightness drawn from the center of the Sun along the direction of
the fixed position angle $PA$ as well as nonradial profiles
$\Delta P(r)$ constructed from a specified point and in a
specified direction. The notations differ in distance $R$, plotted
from the center of the Sun, and in distance $r$, plotted from
another point on the image. All the distances in the plots and
figures are in units of one solar radius ($R_\odot$). In some
cases, difference brightness profiles correspond to the finite
angle average. The averaging had a purpose of improving the
desired signal-to-noise ratio.

\section{Relations at a flat shock discontinuity in the ideal MHD
approximation}

Relations at a discontinuity link various plasma parameters on
either side of the shock discontinuity. They enable us to relate
the Mach number in the medium where the discontinuity moves to the
parameter ratio at the discontinuity. Under the coronal conditions
(at least, at 1-30$R_\odot$ from the center of the Sun), the
Alfv{\'e}n velocity $v_A$ far exceeds the sound velocity $c_S$. We
can therefore take $v_A$ for the typical velocity of disturbance
propagation and find the dependence for the Alfv{\'e}n Mach number
$M_A = u/v_A$, where $u$ is shock wave velocity. An expression for
$M_A$ at the flat shock discontinuity may be obtained in its
analytical form (its derivation is presented in Appendix A).

It is so far impossible to measure magnetic field and temperature
jumps at the shock discontinuity in the corona. But we can
estimate the density ratio $\rho_2/\rho_1$ at the discontinuity
from the white-light coronal brightness (index 1 means that the
value corresponds to the undisturbed region ahead of the shock; 2
to the region behind it). Thus, our prime interest will be with
the $M_A(\rho_2/\rho_1)$ dependence. The angle between the
magnetic field direction ahead of the front and the normal to the
wave front $\theta_{Bn}$, adiabatic index $\gamma$, and the gas to
magnetic pressure ratio $\beta_1$ still remain free parameters in
the equation.

Figure~1 presents sets of $M_A(\rho_2/\rho_1)$ curves for $\gamma
= 2, 5/3$ and 1.46 at different $\theta_{Bn}$ and $\beta_1$. The
left panel in Figure 1 shows the $M_A(\rho_2/\rho_1)$ dependences
for $\beta_1 = 0$ and $\theta_{Bn} = 90^\circ$ and $45^\circ$
(solid and dashed lines respectively). In the right panel Figure
1, $\theta_{Bn} = 90^\circ$ and $\beta_1 = 0$ and 1. The
dependences $M_A(\rho_2/\rho_1)$ at $\gamma = 5/3$ and $\beta_1 =
0$ coincide with similar dependences in \opencite{Kantrowitz1966}
as presented in \opencite{Kennel1984}. Figure 1 implies that the
form of $M_A(\rho_2/\rho_1)$ for quasi-perpendicular shock waves
($\theta_{Bn} > 45^\circ$) depends little on $\theta_{Bn}$ and
$\beta_1$ (at $0 \leq \beta_1 \leq 1$), but varies considerably
depending on $\gamma$.

\begin{figure}
\centerline{\includegraphics[width=0.8\textwidth,clip=]{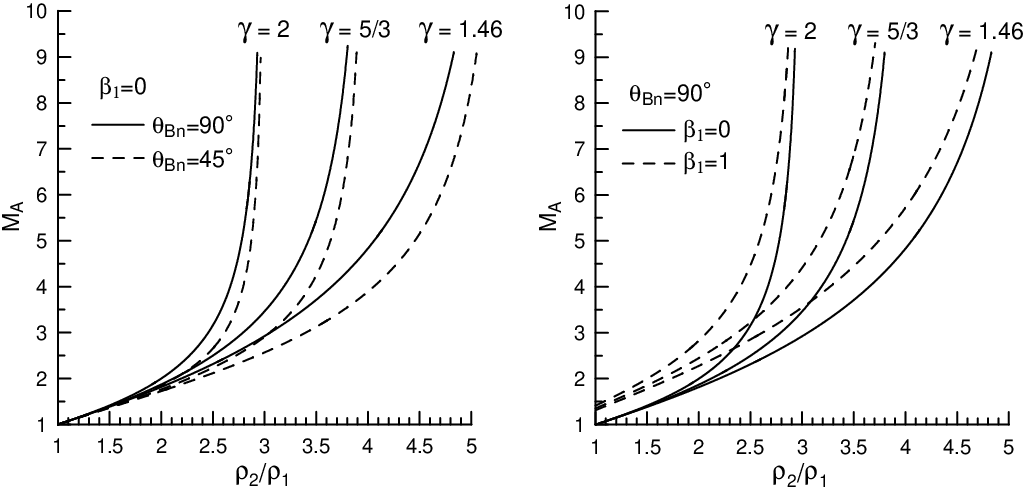}}
\caption{The ideal-MHD calculated dependences $M_A(\rho_2/\rho_1)$
in the flat shock front for various parameters of the upflow
stream.}
\end{figure}

\section{Analysis of experimental data}

\subsection{Event selection criteria}

The main criterion for selecting events for the analysis was a
reliable identification of shock fronts in the events. The
identification was based on the following, previously obtained,
results \cite{Eselevich2010}:
\begin{enumerate}
\item Ahead of CMEs with velocities exceeding $\approx$ 700 km s$^{-1}$, it is
possible to record a shock wave and correctly measure its front
width $\delta_F$ based on Mark 4 and LASCO C2 data. Our
measurements have also shown that the shock front width $\delta_F$
is of order of the free proton path at $1.2R_\odot < R <
6R_\odot$, at least in a limited region located in the direction
of the CME propagation. This provides evidence for the collisional
mechanism of energy dissipation in the shock front at these
distances.
\item A new discontinuity with width $\delta_F^* \ll \delta_F$ is observed to form at $R
> 10R_\odot$ in the frontal part of the shock front. Within the
experimental error, the value of $\delta_F^* \approx$
0.1-0.2$R_\odot$ is independent of $R$, being defined by the LASCO
C3 spatial resolution. This brightness profile transformation from
the front of width $\delta_F$ to the discontinuity of width
$\delta_F^* \ll \delta_F$ was interpreted as transition from a
collisional to a collisionless shock whose front width could not
be resolved in coronal images.
\end{enumerate}
The authors of Paper 1 registered brightness distribution
discontinuities interpreted as shock waves, at the front of 13
selected CMEs moving at over 1500 km s$^{-1}$. Since the
heliocentric distances at which the discontinuities were observed
generally exceeded 6-10$R_\odot$, and the minimum width of these
discontinuities did not exceed 0.1-0.2$R_\odot$, they appear to be
associated with collisionless shock waves.

The flat shock wave condition $\delta_F^* \ll R_F$ ($R_F$ is the
radius of wavefront curvature) is better satisfied at large
distances because it is there that fronts of width
0.1-0.2$R_\odot$ are encountered. Moreover, the shock wave
strength is sufficiently great ($\rho_2/\rho_1 \geq 2$) at these
distances; i.e. the shock wave has already been formed and is
stable. Therefore, we chose for our analysis those instances of
time when the shock front was farther than 10$R_\odot$.

Comparing the experimental data with the calculations of relations
in the shock front requires absolute measurements of density; it
is therefore very important to select events with minimised
influence of factors affecting the measurements. We have chosen
ten CMEs having shock waves in front, while keeping in mind the
following requirements:

\begin{enumerate}
\item The origin of the CMEs was near the limb and they propagated in
the plane of the sky; i.e., their measured positions and
velocities were close to the true values.
\item 12-24 hours before each of the selected events, there were no
other CMEs capable of noticeably changing the undisturbed solar
wind conditions in the coronal region under study.
\end{enumerate}

These events are listed in Table~1. The third column in the Table
lists the coordinates of the CME-associated source region on the
disk. The coordinates for the first two events were borrowed from
\opencite{Cremades2004}, for the others from the catalogue
\opencite{Yashiro2004}. Since the origin of all the ten CMEs was
near the limb, let us call them `limb CMEs'. According to the
classification accepted in the catalogue \opencite{Yashiro2004},
they include halo (events 3, 4, 7, 8 in Table~1) and partial halo
CMEs (event 2). Let us not confuse limb CMEs with coronal
ejections appearing at longitudes below 60-70$^\circ$, which are
therefore no limb CMEs.

\setlength{\tabcolsep}{5pt}

\begin{table}
\caption{  } 

\begin{tabular}{ccccccccc}

  \hline
1 & 2 & 3 & 4 & 5 & 6 & 7 & 8 & 9  \\ \hline

Event & Time & Source & $R$ & $\delta P$ & $V$ & $M_A$ & $\delta
N$ & $\rho_2/\rho_1$ \\

  & & coord. & ($R_\odot$) & ($P_{MSB}$) & (km s$^{-1}$) &  & (cm$^{-3}$) & \\
 \hline

1 &  \multicolumn{8}{c}{1997 September 20} \\
  & 15:26:47 & S24W102 & 23.3 & $2.5\times10^{-13}$ & 730 & 1.5 & $2.4\times10^3$ & 2.5 \\
  & 16:01:23 &      & 25.4 & $1.7\times10^{-13}$ & 710 & 1.5 & $1.9\times10^3$ & 2.5 \\
  \hline

2 &  \multicolumn{8}{c}{1998 April 20} \\
  & 11:48:11 & S40W106 & 17.0 & $1.1\times10^{-12}$ & 1630 & 3.7 & $5.6\times10^3$ & 2.8 \\
  & 12:38:07 &       & 24.0 & $3.5\times10^{-13}$ & 1640 & 4.7 & $3.6\times10^3$ & 3.4 \\
  \hline

3 &  \multicolumn{8}{c}{1999 July 25} \\
  & 14:40:10 & N23W81  & 11.0 & $2.4\times10^{-12}$ & 1280 & 2.1 & $5.1\times10^3$ & 1.6 \\
  & 15:16:11 &       & 15.0 & $8.0\times10^{-13}$   & 1230 & 2.4 & $3.2\times10^3$ & 1.8 \\
  & 16:16:17 &       & 21.0 & $3.2\times10^{-13}$ & 1070 & 2.5 & $2.5\times10^3$ & 2.2 \\
  & 16:40:25 &       & 23.0 & $2.3\times10^{-13}$ & 970  & 2.3 & $2.2\times10^3$ & 2.3 \\
  \hline

4 &  \multicolumn{8}{c}{2002 April 21} \\
  & 03:16:57 & S16W84  & 26.1 & $2.9\times10^{-13}$ & 2422 & 7.9 & $3.5\times10^3$ & 3.7 \\
  \hline

5 &  \multicolumn{8}{c}{2002 August 16} \\
  & 07:39:53 & N07W83  & 13.5 & $1.7\times10^{-12}$ & 1340 & 2.5 & $5.5\times10^3$ & 2.1 \\
  & 08:15:51 &       & 17.6 & $9.0\times10^{-13}$ & 1490 & 3.4 & $4.9\times10^3$ & 2.7 \\
  & 08:39:50 &       & 21.0 & $3.7\times10^{-13}$ & 1390 & 3.5 & $2.9\times10^3$ & 2.4 \\
  & 09:15:50 &       & 24.5 & $2.0\times10^{-13}$ & 1130 & 3.0 & $2.1\times10^3$ & 2.5 \\
  \hline

6 &  \multicolumn{8}{c}{2003 November 4} \\
  & 13:41:57 & N08W90  & 12.2 & $1.5\times10^{-12}$ & 1000 & 1.6 & $3.9\times10^3$ & 1.6 \\
  & 14:17:55 &       & 15.4 & $6.0\times10^{-13}$ & 980  & 1.8 & $2.5\times10^3$ & 1.6 \\
  & 14:41:47 &       & 17.3 & $7.7\times10^{-13}$ & 960  & 1.9 & $4.1\times10^3$ & 2.3 \\
  & 15:17:27 &       & 20.3 & $4.0\times10^{-13}$ & 980  & 2.2 & $2.9\times10^3$ & 2.4 \\
  \hline

7 &  \multicolumn{8}{c}{2005 August 23} \\
  & 15:40:45 & S14W90  & 10.6 & $1.5\times10^{-12}$ & 1490 & 2.5 & $3.0\times10^3$ & 1.3 \\
  & 16:16:53 &       & 15.1 & $3.5\times10^{-13}$ & 2070 & 4.5 & $1.4\times10^3$ & 1.3 \\
  & 16:40:42 &       & 20.6 & $4.3\times10^{-13}$ & 2120 & 5.7 & $3.2\times10^3$ & 2.5 \\
  & 17:16:50 &       & 25.4 & $1.8\times10^{-13}$ & 1550 & 4.6 & $2.1\times10^3$ & 2.5 \\
  \hline

8 &  \multicolumn{8}{c}{2005 September 5} \\
  & 10:41:15 & S07E81  & 12.2 & $8.0\times10^{-13}$ & 2070 & 3.9 & $2.1\times10^3$ & 1.3 \\
  & 11:17:24 &       & 18.8 & $6.0\times10^{-13}$ & 2360 & 6.0 & $3.8\times10^3$ & 2.5 \\
  & 11:41:16 &       & 24.1 & $3.5\times10^{-13}$ & 2590 & 8.1 & $3.6\times10^3$ & 3.4 \\
  \hline

9 & \multicolumn{8}{c}{2007 December 31} \\
  & 02:39:56 & S08E81  & 11.4 & $1.1\times10^{-12}$ & 1150 & 1.9 & $2.5\times10^3$ & 1.3 \\
  & 03:39:53 &       & 17.5 & $5.0\times10^{-13}$ & 1100 & 2.3 & $2.7\times10^3$ & 1.9 \\
  & 04:40:06 &       & 22.7 & $2.7\times10^{-13}$ & 1000 & 2.4 & $2.5\times10^3$ & 2.4 \\
  \hline

10 & \multicolumn{8}{c}{2008 March 25} \\
  & 21:16:29 & S13E78  & 15.2 & $6.5\times10^{-13}$ & 1040 & 2.0 & $2.7\times10^3$ & 1.7 \\
  & 21:40:06 &       & 17.2 & $3.8\times10^{-13}$ & 990  & 2.0 & $2.0\times10^3$ & 1.6 \\
  \hline

\end{tabular}
\end{table}

\subsection{Registering shock fronts ahead of CMEs}

The shape of the CME outer boundary (and the adjacent shock wave)
is usually nearly circular when the CME velocity exceeds the
critical value $u_C$ (Figure~2). The center of this circle (point
O in the figure) may be defined by distance $R_C$ from the center
of the Sun (point S) and on the position angle $PA_C$. The
difference brightness profiles $\Delta P(r)$ used to determine
shock wave position were drawn from the center of the circle. The
position of each profile was given by angle $\alpha$ to the
direction of CME propagation. This angle is positive
counterclockwise.

\begin{figure}
\centerline{\includegraphics[width=0.5\textwidth,clip=]{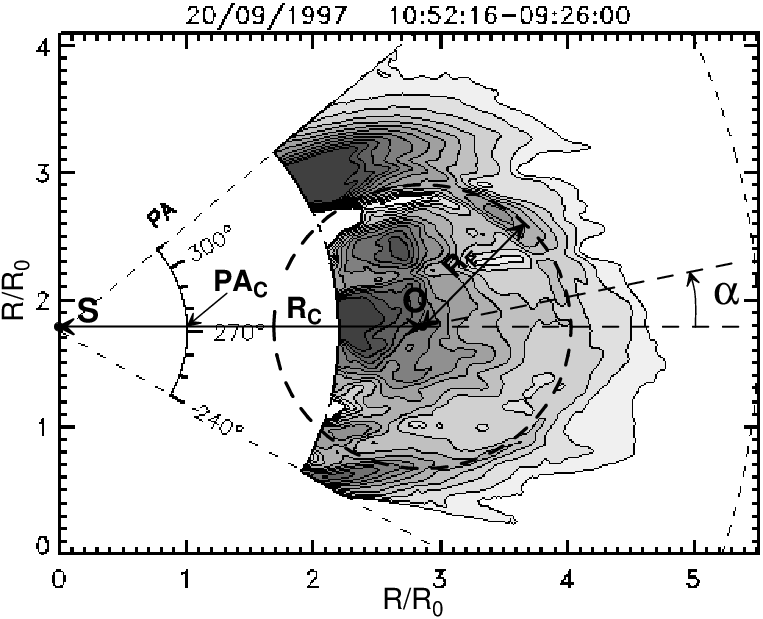}}
\caption{An example of a CME moving at a supersonic velocity
(difference brightness isolines, LASCO C2 data). This figure
includes the notations employed for referencing and constructing
difference brightness profiles required for measurements in the
shock front. The relevant explanations are given in the text. }
\end{figure}

A shock associated discontinuity in difference brightness
distributions is generally observed in a limited region in the
direction of CME propagation \cite{Eselevich2010}. The value of
$\alpha$ at which shocks were recorded did not exceed $\pm
10^\circ$ in all the ten selected events. For this range of angles
$\alpha$, the distances from the center of the Sun, $R$, and from
the CME center, $r$, to the shock front may be related by a simple
relation $R \approx R_C + r$, with an accuracy of 0.1$R_\odot$.
However, even a-few-degree change in $\alpha$ and a-few-pixel
change in the CME center position significantly affect the
discontinuity width as recorded in difference brightness profiles.
An accurate choice of the O center position almost always provides
the minimum discontinuity width
comparable to the LASCO C3 spatial resolution.

Difference brightness profiles $\Delta P(r)$ drawn from the CME
center were used to register shock fronts and measure their
positions and amplitudes. Figures 3-6 present examples of such
distributions for four events in Table 1. Each of the figures
corresponds to one of the events, and each plot in the figures
corresponds to a certain instant of time. The time difference used
to construct the difference brightness, the CME center position
(angle $PA_C$ and distance $R_C$), and the direction along which
the profile was drawn (angle $\alpha$) are given for each plot. In
all the cases, we used averaging in the range of angles $\delta
\alpha = 2^\circ$. For convenience, each of the plots is
reconstructed depending on the distance $R$ from the center of the
Sun, though the scanning direction may slightly differ from the
radial direction. Each plot presents two distributions: 1) for the
instants when CMEs and shocks were recorded and 2) the difference
brightness distributions for two neighboring instants before the
CMEs appeared in the C3 field of view (the distributions are
denoted by black and light circles respectively). The latter make
it possible to estimate the noise level; in particular, they were
used to find the error ($1\sigma$) in measuring the shock wave
amplitude $\delta P$. At large distances, the noise level is
comparable with the amplitude in some of the events.

\begin{figure}
\centerline{\includegraphics[width=0.5\textwidth,clip=]{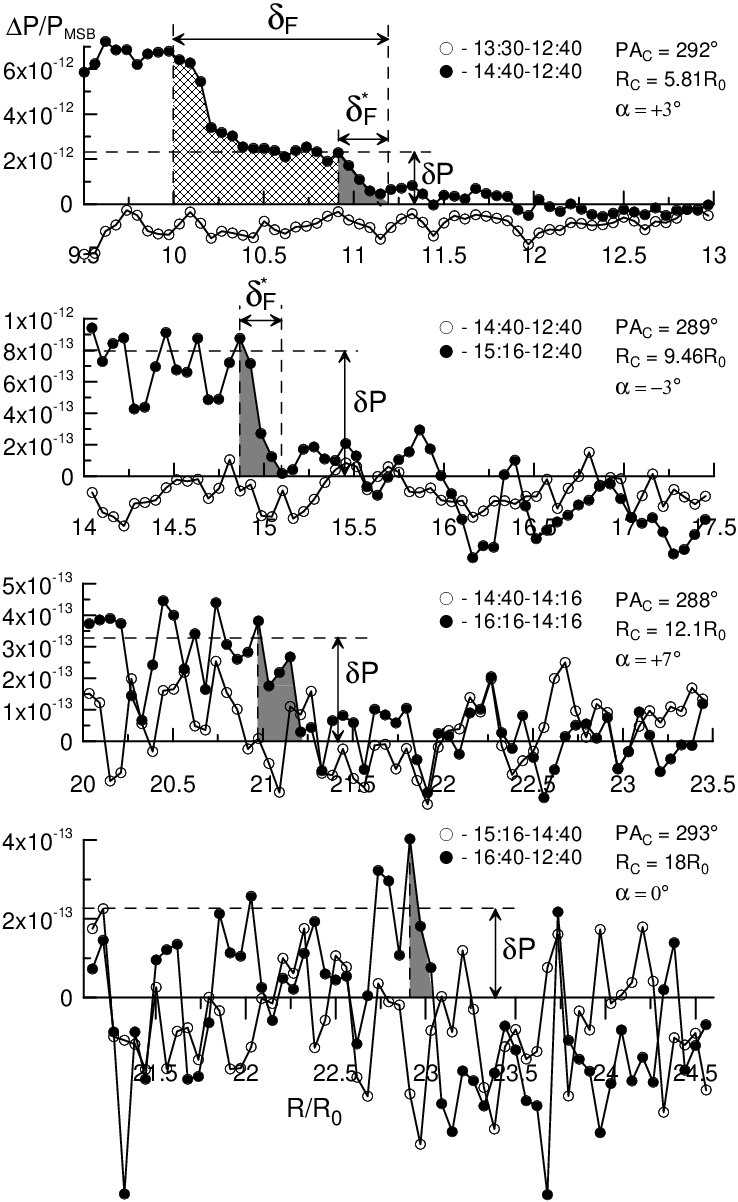}}
\caption{Black circles mark the difference brightness $\Delta
P(R)$ distributions at successive instants of time for the 1999
July 25 CME. Light circles show the $\Delta P(R)$ distributions
constructed for the instant of time just before the CME appears in
the C3 field of view.}
\end{figure}

\begin{figure}
\centerline{\includegraphics[width=0.5\textwidth,clip=]{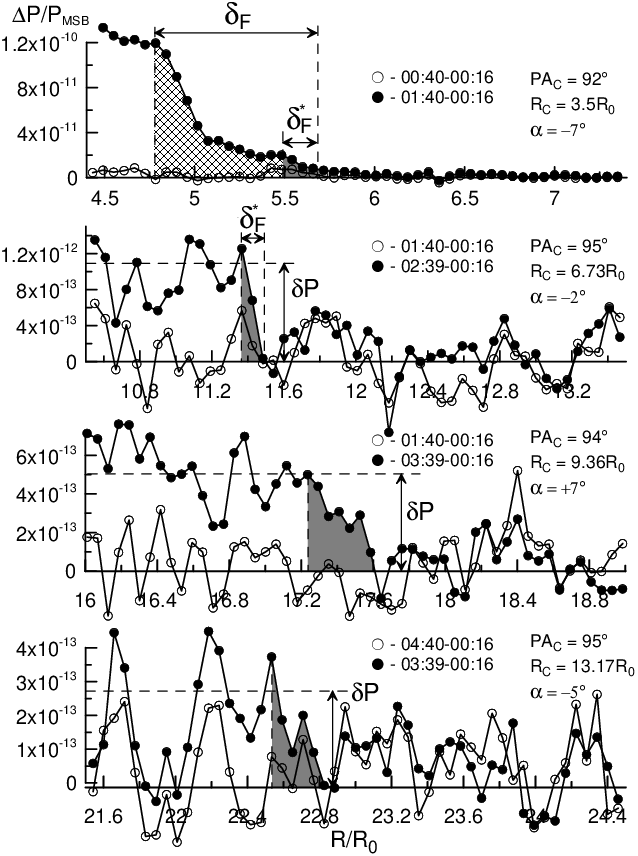}}
\caption{The same as in Fig. 3, but for the CME of 31 December
2007.}
\end{figure}

\begin{figure}
\centerline{\includegraphics[width=0.5\textwidth,clip=]{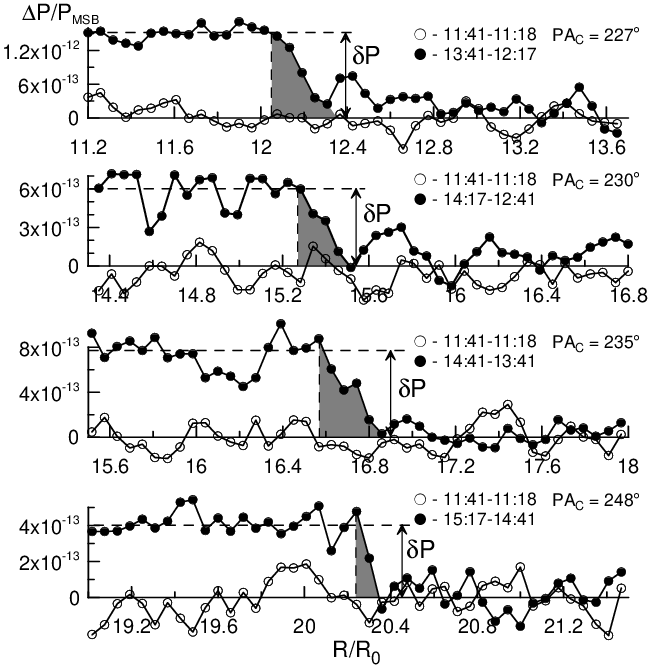}}
\caption{The same as in Fig. 3, but for the CME of 4 November
2003.}
\end{figure}

\begin{figure}
\centerline{\includegraphics[width=0.5\textwidth,clip=]{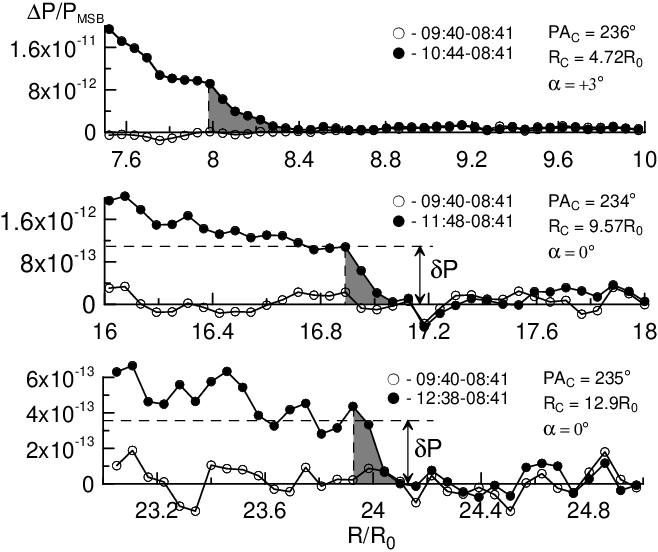}}
\caption{The same as in Fig. 3, but for the CME of 20 April 1998.}
\end{figure}

All the events have common features, but differ in some respects.
The differences require some commentary. The upper plots in
Figures 3 and 4 show the stage when a collisional shock wave with
front width $\delta_F$ transforms into a collisionless
discontinuity with front width $\delta_F^* \ll \delta_F$, for the
1999 July 25 and 2007 December 31 events, respectively. Such a
transformation usually takes place at $R < 10R_\odot$. The
location and amplitude of $\delta P$ were determined only for
shock discontinuities with front width $\delta_F^*$ at $>
10R_\odot$. Another event occurred in the same coronal region
approximately 12 hours before the CME, during the 1999 July 25
event (Figure~3). As a result, the undisturbed difference
brightness constructed for a slightly earlier instant of time
(light circles in the two upper plots) has a mean value that is
significantly different from zero. In determining $\delta P$, we
used averaging over a region behind the front whenever possible
(Figures 3-6). The front width $\delta_F^*$ slightly varies but is
almost constant at $> $10-15$R_\odot$ (the minimum value is
0.1-0.2$R_\odot$), and the observed variations may have been due
to noise.

The plots in Figures 3-6 as well as similar plots for other events
were used to determine the shock wave distance from the center of
the Sun (Column 4 in Table 1) and shock wave amplitude (Column 5)
at a given instant of time. The obtained distances were employed
to calculate shock wave velocities (Column 6). The velocity
calculations took into account information on shock wave locations
at $< 10R_\odot$ at earlier instants of time, not included in
Table 1. The velocity at a given instant of time is the average
velocity for the intervals before and after this instant, except
for the last instant, for which the velocity is simply the
velocity as determined for the previous interval.

\subsection{Estimating shock wave strength and comparing the
results with MHD calculations}

\subsubsection{Experimental values for estimating the shock wave
strength}

Our aim is to compare the experimental dependence
$M_A(\rho_2/\rho_1)$ with the analogous dependence derived from
MHD calculations. The shock wave strength is defined as
$\rho_2/\rho_1 \approx 1 + \delta N/N_0$, where $\delta N$ is the
absolute jump in electron density in the shock front, $N_0$ is the
undisturbed plasma electron density immediately ahead of the shock
front. To find $M_A$, we have to know the local Alfv{\'e}n
velocity ($V_A$) and the velocity ($V_{SW}$) of the solar wind
relative to which the shock waves move.

Since $N_0$, $V_A$ and $V_{SW}$ cannot be determined directly from
the available experimental data, let us use their mean values
typical of the quasi-stationary slow solar wind (SW) flowing in
the streamer belt. The distance $R$ dependences of $V_{SW}$,
$V_A$, and $N_0$ (Figure~7) were taken respectively from
\cite{Wang2000}, \cite{Mann1999}, and \cite{Saito1977}. The
validity of using the above dependences is obvious enough because
the shock front parts under analysis are located near the axis of
the CMEs propagating in the slow SW. This is also confirmed by the
successful application of the dependences $V_{SW}(R)$ and $V_A(R)$
in the analysis in (\opencite{Eselevich2008}; see Figure~3) and,
to some extent, of the dependence $N_0(R)$ in Paper 1. Applying
$N_0(R)$, however, has a limitation mentioned above -- there
should be no other CMEs capable of changing the undisturbed SW
conditions considerably in the region under study in the previous
24 hours. Column 7 in Table 1 gives calculated values of $M_A =
(V-V_{SW})/V_A$, where $V_{SW}$ and $V_A$ values derived from the
above experimental dependences.

\begin{figure}
\centerline{\includegraphics[width=0.5\textwidth,clip=]{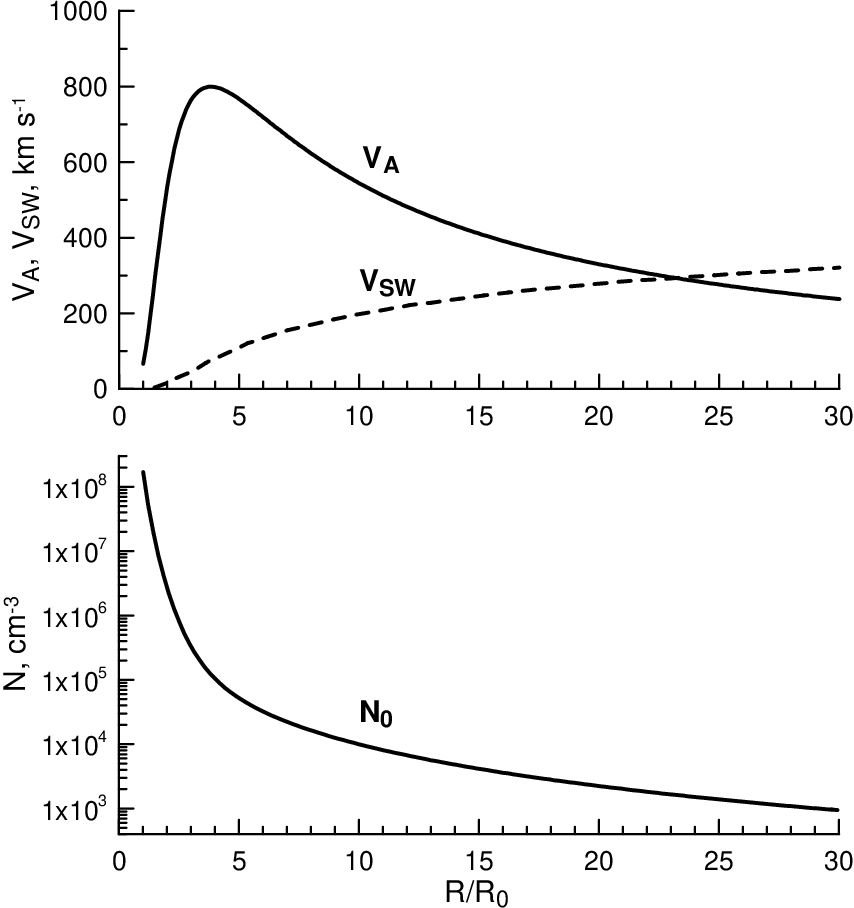}}
\caption{The experimental distributions employed to calculate
$M_A$ and $\rho_2/\rho_1$. The upper panel: the solid curve is the
Alfv{\'e}n velocity $V_A(R)$ in \protect\opencite{Mann1999}, the
dashed curve is the slow SW velocity $V_{SW}(R)$ in
\protect\opencite{Wang2000}; the lower panel is the electron
density in the corona in \protect\opencite{Saito1977}.}
\end{figure}

The density jump $\delta N$ in the shock front remains an unknown
value and is to be determined. It is proportional to brightness
jump $\delta $ in the difference brightness distributions
(Figures 2-6). The relationship between $\delta $ and $\delta N$
depends on the unknown plasma distribution in the front. However,
by specifying mean line-of-sight size $l$ of the shock front, it
is possible to calculate the mean density jump from the measured
brightness discontinuity, using known relations
\cite{Billing1966}. As long as we are dealing here with limb
events, we may assume that the shock front is in the plane of the
sky.

In Paper 1, the mean size of the shock front -- $l = 1R_\odot$ --
was used to calculate $\delta N$ from $\delta P$ for all the cases
of the nonlimb CMEs. In what follows we will try to justify
quantitatively a somewhat different value of $l$ for the limb CMEs
in Table 1, common for all the events under investigation as well.

\subsubsection{Estimating the ultimate compression in a
collisionless shock wave front}

In the front of a quasi-perpendicular ($\theta_{Bn} > 45^\circ$)
collisionless shock wave, heating of ions can occur in a plane
perpendicular to the magnetic field direction \cite{Zimbardo2009}.
This transverse motion is two-dimensional, which corresponds to
two degrees of freedom ($i = 2$) and consequently to the adiabatic
index $\gamma = (i+2)/i = 2$. In the case of isotropic turbulence
in the shock front, the number of degrees of freedom will tend to
$i = 3$, and the adiabatic index to $\gamma = 5/3$
\cite{Sagdeev1966}. It is therefore reasonable to expect that the
adiabatic index $\gamma$ will be between 5/3 and 2 in the front of
a quasi-perpendicular collisionless shock. This means that maximum
$\rho_2/\rho_1$ values at large Mach numbers must not exceed 4
(see Figure~1 or Appendix).

This assumption is supported by the results of direct measurements
in the Earth's bow shock and interplanetary shocks. These
measurements do not produce $\rho_2/\rho_1 > 4$ even for the
largest Mach numbers. To confirm this, the upper panel of Figure~8
presents the experimental dependences $M_A(\rho_2/\rho_1)$ for bow
(solid marks) and interplanetary (light marks) shocks based on
data of various authors. The values of $\rho_2/\rho_1$ in these
dependences were taken from \cite{Zastenker1983} and
\cite{Formisano1973,Greenstadt1980,Bame1979,Bale2005} for
interplanetary and bow shocks respectively. In some cases, the
calculation of the Alfv{\'e}n Mach number was based on
hour-averaged OMNI data (\url{http://omniweb.gsfc.nasa.gov/}) on
the magnetic field, SW density and velocity (for the bow shock).
All the obtained $\rho_2/\rho_1$ do not exceed 4. This is also
consistent with the $M_{MS}(\rho_2/\rho_1)$ plot in Figure 6 in
\opencite{Formisano1973}. That figure implies $\rho_2/\rho_1 \leq
4$ for all 42 crossings of the bow shock with large magnetosonic
Mach numbers $M_{MS} = 4-12$ (by definition, $M_A \geq M_{MS}$).

\begin{figure}
\centerline{\includegraphics[width=0.6\textwidth,clip=]{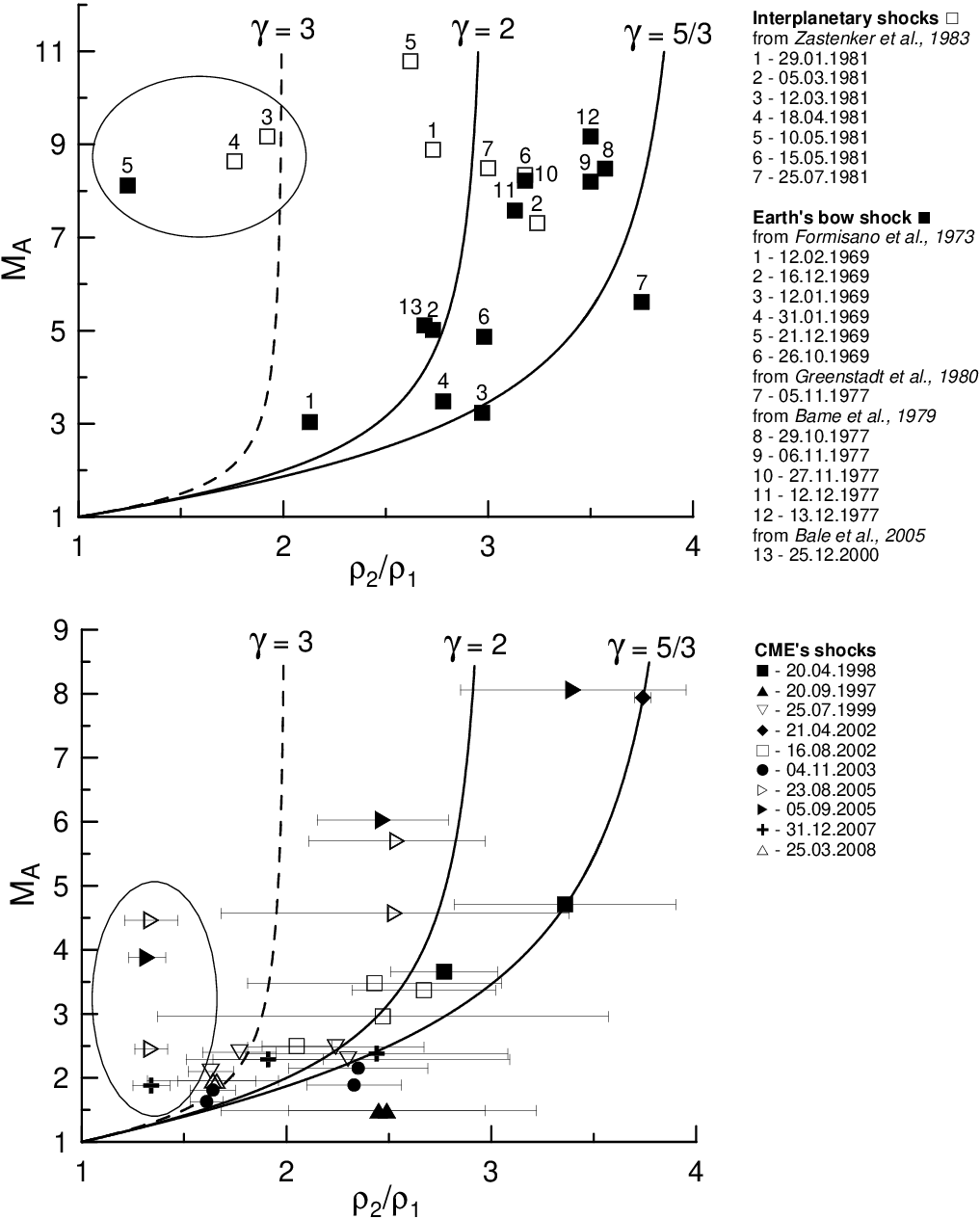}}
\caption{The experimental dependence $M_A(\rho_2/\rho_1)$: the
upper panel is the bow shock (dark squares) and interplanetary
shocks near the Earth's orbit (light squares); the lower panel
displays ten shocks associated with limb CMEs. The solid and
dashed curves denote ideal MHD calculations for perpendicular
($\theta_{Bn} = 90^\circ$) shocks at $\gamma = 5/3$, 2 and 3. }
\end{figure}

The upper panel of Figure~8 implies that the experimental points
are mainly located near the calculated curves for $\gamma = 5/3$
and $\gamma = 2$, i.e. in the collisionless shock front $\gamma
\geq 5/3$. Note that the effective adiabatic index $\gamma$ is
smaller in the undisturbed SW. For example, \opencite{Totten1995}
derived the empirical, mean adiabatic index for protons -- $\gamma
= 1.46 \pm 0.04 < 5/3$ -- using Helios 1 data for the SW moving at
$V_{SW} = 300-800$ km s$^{-1}$. The fact that shocks are
characterized by greater values of $\gamma \approx 5/3-2$ must be
a result of collisionless processes inside the shock front.

\subsubsection{Determining the effective line-of-sight size $l$ of the shock
front}

By analogy with the heliosphere it is reasonable to expect that
the adiabatic index $\gamma \geq 5/3$, and maximum density ratio
is less than 4 for a collisionless shock in the solar corona. Let
this conclusion serve as a basis for estimating $l$. For the
purpose, we will examine the 2002 April 21 CME in more detail
(number 4 in Table 1). This CME had the highest velocity of the
ten selected events, and a maximum compression was observed in the
shock front in front of the CME at about 26$R_\odot$ from the
center of the Sun at $t$=03:16:57. To enhance the accuracy, we
constructed difference brightness distributions for several angles
$\alpha$ in the range $-20^\circ$ to $+20^\circ$. Five of the
distributions are shown in Figure~9. Besides, to determine the
brightness jump $\delta P$ the brightness profile was averaged on
either side of the discontinuity. The event was accompanied by an
intense flux of energetic particles which manifested themselves as
bright dots and scratches on the coronal images. In some cases,
filtration failed to completely remove such noise from the images.
However, it can easily be detected in the brightness profiles (one
of the examples is given in the upper plot of Figure~9). These
parts of the brightness profiles were not used to determine
$\delta P$.

\begin{figure}
\centerline{\includegraphics[width=0.7\textwidth,clip=]{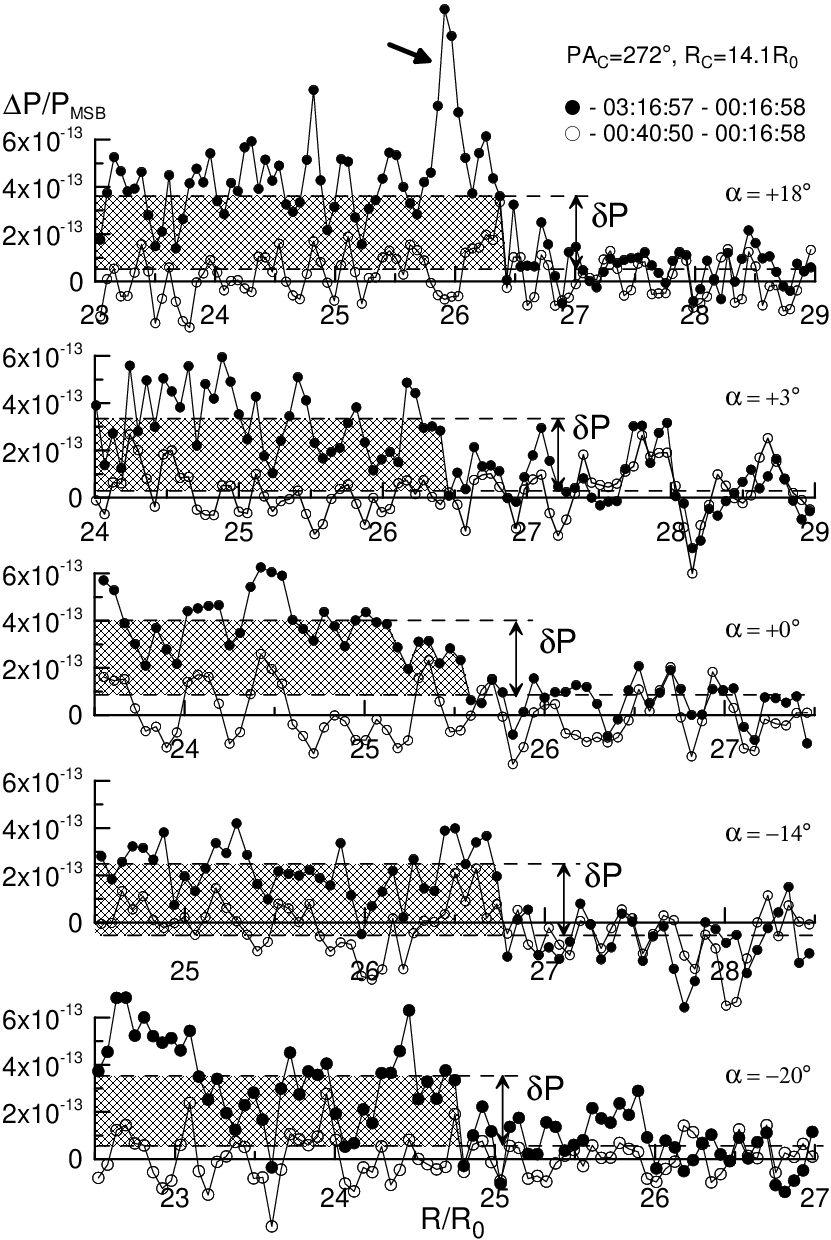}}
\caption{The difference brightness $\Delta P(R)$ distributions for
several angles $\alpha$ for the CME of 21 April 2002; the shock
front ahead of this CME had the highest compression of all the
events under study. }
\end{figure}

Table 2 summarizes the measured $\delta P$ values for various
angles $\alpha$ and the distances at which they were observed. The
distances differ slightly, indicating that the shock front is
inhomogeneous along the direction of angle $\alpha$. Determination
of the front velocity and density discontinuity employed their
mean values. The mean brightness discontinuity was $\delta P =
2.9\times 10^{-13} P_{MSB}$, and the mean velocity $V = 2422$ km
s$^{-1}$. This velocity corresponded to $M_A = 7.94$ at the
average distance $R = 26.1R_\odot$, where the shock front was
recorded. The undisturbed density at the distance was $N_0 =
1.3\times 10^3$ cm$^{-3}$.

\begin{table}
\caption{  } 

\begin{tabular}{ccc}

\hline

$\alpha$, deg. & $R$, $R_\odot$ & $\delta P$,
$\times10^{-13}P_{MSB}$
\\ \hline
-20 & 24.8 &   2.78 \\

-17 & 26.0 &   2.86 \\

-14 & 26.7 & 3.02 \\

-11 & 26.9 & 2.00 \\

-1  & 25.4  &  3.24 \\

0   & 25.5  &  3.11 \\

+3  & 26.4  &  2.57 \\

+6  & 25.7  & 3.37 \\

+10 & 26.8  & 2.94 \\

+18 & 26.4  & 3.36 \\

+19 & 26.3 & 2.65 \\

\hline

&  $26.1 \pm 0.66$ & $2.9 \pm 0.4$ \\

\hline

\end{tabular}
\end{table}

The brightness jump $\delta P$ at a given distance $R$ allows us
to calculate $\delta N$ and, taking $N_0$ into account, derive
$\rho_2/\rho_1$ from it. As was stated above, we must set the
shock front size $l$ along the line of sight for the purpose. The
curve in Figure~10 demonstrates how the calculated $\rho_2/\rho_1$
varies with the selected size $l$ (in units of solar radius) under
the conditions of the event under study ($\delta P = 2.9\times
10^{-13} P_{MSB}$, $R = 26.1R_\odot$, $N_0 = 1.3\times 10^3$
cm$^{-3}$). The main critical parameter is the adiabatic index
$\gamma$ because the $\beta$ and $\theta_{Bn}$ dependences are
rather weak. The range of maximum $\rho_2/\rho_1$ (i.e. as
$M_A\rightarrow\infty$) for $\gamma$ between 5/3 and 2 is
indicated by hatching in Figure~10.

\begin{figure}
\centerline{\includegraphics[width=0.5\textwidth,clip=]{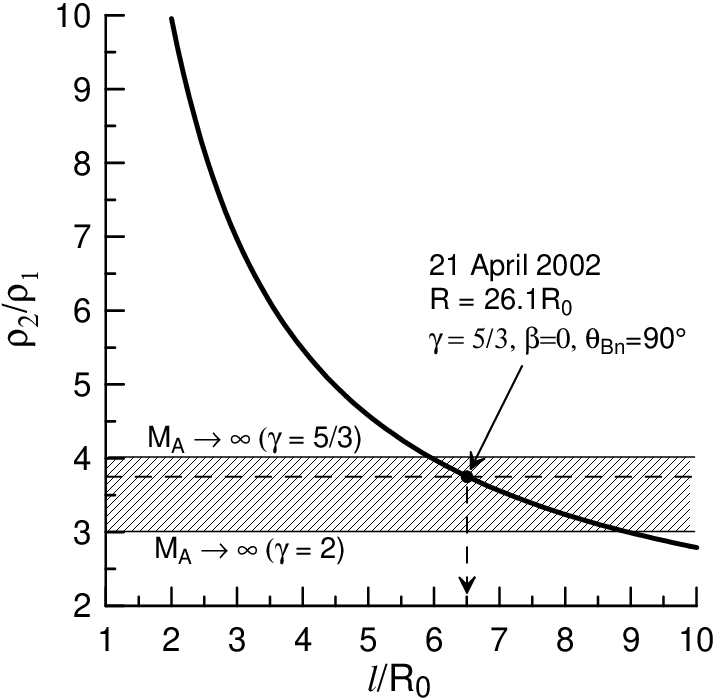}}
\caption{The $\rho_2/\rho_1$ values depending on the effective
length $l$ along the line of sight. They were calculated for the
shock wave parameters at 03:16:57 ahead of the CME of 21 April
2002.}
\end{figure}

Let us assume that an extreme case is observed in the event:
$\gamma = 5/3$, $\beta = 0$ and $\theta_{Bn} = 90^\circ$, i.e.
there is a perpendicular shock with maximum compression. Then, the
MHD calculations imply that the shock wave strength $\rho_2/\rho_1
= 3.74$, being close to the limiting value of 4. The plot in
Figure~10 indicates that this strength corresponds to $l =
6.5R_\odot$. This size was employed to calculate $\delta N$ and
$\rho_2/\rho_1$ in all the ten selected events, their values are
listed in columns 8 and 9 of Table 1.

\subsubsection{Experimental dependence $M_A(\rho_2/\rho_1)$ for shocks in the
corona}

The calculated values in the lower panel of Figure~8 are used to
compare the experimental dependence $M_A(\rho_2/\rho_1)$ for ten
shocks with calculation results for perpendicular shocks
($\theta_{Bn} = 90^\circ$) at $\gamma = 5/3$ and $\gamma = 2$
(solid lines). The error bar due to the error in determining the
brightness jump is indicated for the experimental points.

The dependences $M_A(\rho_2/\rho_1)$ suggest that as a whole they
fit the calculated curves reasonably well in spite of the
considerable scatter of experimental points. This allows the
following conclusions to be made:
\begin{enumerate}
\item The choice of the mean value $l = 6.5R_\odot$ has proved to be reasonable
enough despite the fact that the shock fronts in question were
observed at various distances ($10R_\odot < R < 30R_\odot$) and
had velocities $V\approx$ 800-2500 km s$^{-1}$.
\item The comparison results agree with the assumption that the
effective adiabatic index $\gamma$, characterizing the processes
within the front, is largely between 2 and 5/3, which corresponds
to the effective number of degrees of freedom of motion being 2 to
3. This also testifies that the structures under investigation are
collisionless shock fronts.
\end{enumerate}

Of special notice are several outstanding points which correspond
to shocks with sufficiently large Mach numbers but with small
$\rho_2/\rho_1$ (outlined in upper and lower panels of Figure~8).
They all lie to the left of the calculated curve with $\gamma =
3$. This corresponds to the effective number of degrees of freedom
< 1. If these points do not result from errors, we have to admit
that in these cases the shock is not in a stationary state
(physically justified dependence $M_A(\rho_2/\rho_1)$ becomes
invalid for the points). It may result from various reasons. For
example, in the case of the bow shock, the nonstationarity of the
front may result from a change in the parameters of the upflow
undisturbed SW.

The experimental results in Table 1 allow certain conclusions to
be made with regard to how the shock front propagates in a coronal
plasma where concentration rapidly decreases with distance. On the
example of three different events in Figure~11, one can see that
the shock wave strength and Alfv{\'e}n Mach number, on the
average, either change insignificantly or increase with distance
at distances up to $R \leq 30R_\odot$. Such behaviour is also
typical of other events.

\begin{figure}
\centerline{\includegraphics[width=0.5\textwidth,clip=]{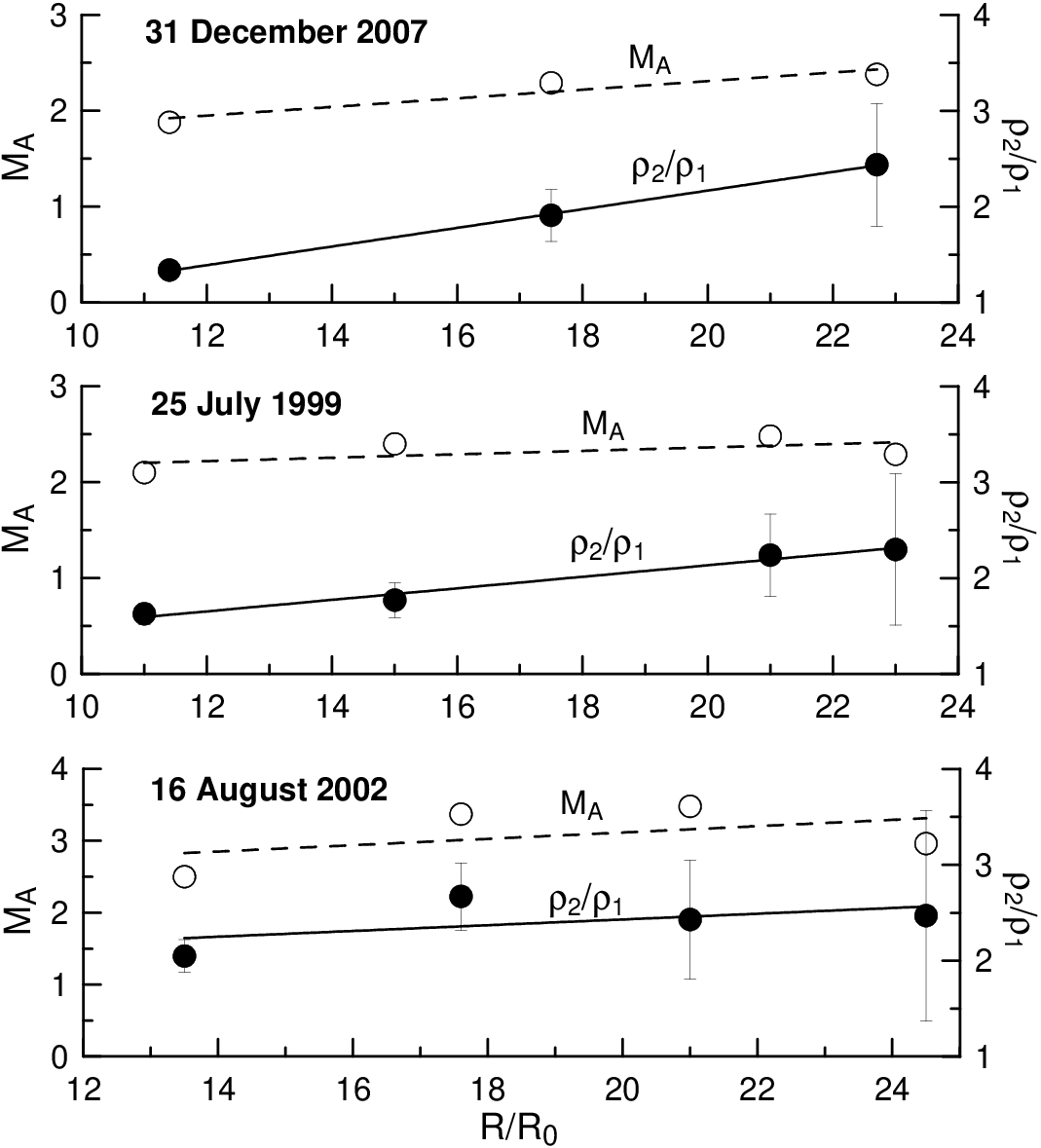}}
\caption{$Œ_€$ (light circles) and $\rho_2/\rho_1$ variations
(dark circles) with distance in the events on 31 December 2007, 25
July 1999, and 16 August 2002.}
\end{figure}

\section{Comparison with results of Paper 1}

Density ratios $\rho_2/\rho_1$ (marked $\Gamma_{CR}$) have been
found in Paper 1, for eleven shocks driven by CMEs moving at
1500-2000 km s$^{-1}$. The parameters of the events are in Tables
1 and 2 in Paper 1. They are numbered 1, 2, 4, 6, 7, 8, 9, 10, 11,
12, 14. The table velocity and distance values were used to
calculate $M_A$ values. The asterisks in Figure~12 mark the values
corresponding to the events on the $M_A(\rho_2/\rho_1)$ plot.
Since the $\rho_2/\rho_1)$ values were derived in Paper 1 for $l =
1R_\odot$, the density ratios were recalculated for $l =
6.5R_\odot$ to compare them with our findings. The recalculated
values are marked by dark circles in Figure~12. The light circles
indicate our values (the same as in the lower panel of Figure~8).

\begin{figure}
\centerline{\includegraphics[width=0.5\textwidth,clip=]{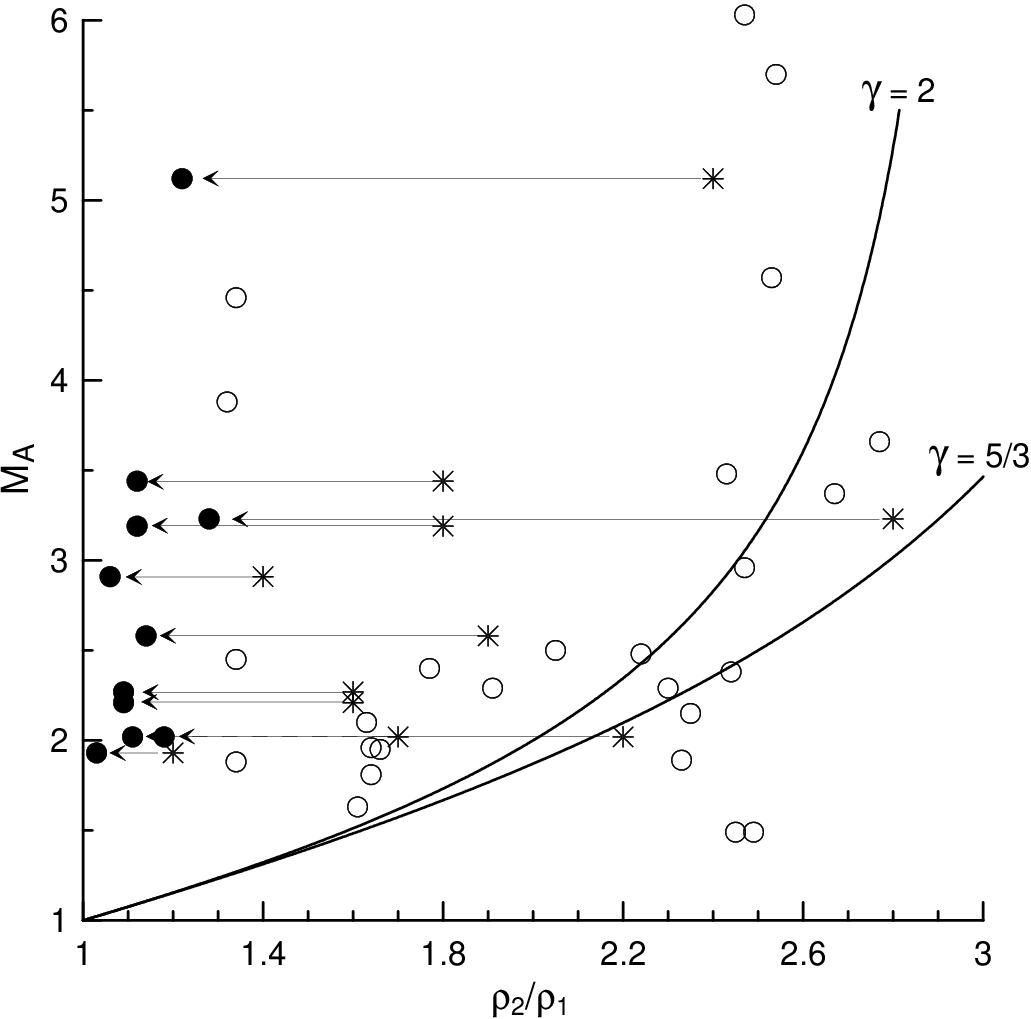}}
\caption{The experimental dependence $M_A(\rho_2/\rho_1)$
constructed from the $\rho_2/\rho_1)$ values derived in this paper
(light circles) and in Paper 1 for $l = 1R_\odot$ (asterisks) and
recalculated for $l = 6.5R_\odot$ (dark circles).}
\end{figure}

It is obvious that there is a significant difference between
findings in this paper and Paper 1: the density ratios in Paper 1
are much smaller than the values we obtain here. Note that the
method for determining $\rho_2/\rho_1$ is identical in both
papers. The only difference is the value of $l$: $1R_\odot$ in
Paper 1 and $6.5R_\odot$ in our paper. There may be several
reasons for the difference. We list them in the order of their
influence on the result:
\begin{enumerate}
\item presence of other CMEs capable of markedly changing the
undisturbed solar wind density in the coronal region under study,
less than twelve hours before the CME (events 1, 2, 4, 8, 10, 12
in Paper 1);
\item a non-stationary shock front, when the registration point is at $R
< $10-15$R_\odot$, and $\rho_2/\rho_1 < 2.0-2.5$ (events 1, 4, 6,
8, 9, 10, 11, 12, 14);
\item
the CME source is rather far from the limb (nonlimb event)
resulting in a noticeable error in determining distance $R$ at
which the shock is registered (and thus $\rho_1$ and $V_A$) and
its velocity $V$. This may also result in decreased intensity of
scattered light and, thus, decreased measured value of $\delta $
(events 1, 2, 6, 8, 9, 10, 11, 12).
\end{enumerate}

Finally, the last comment. Paper 1 does not include $\rho_2/\rho_1
> 2.8$. But if we set $l = 1R_\odot$ for our events with $\rho_2/\rho_1 > 2.5$ at $l
= 6.5R_\odot$, we will obtain $\rho_2/\rho_1 > 10$. Obviously,
this value far exceeds the maximum values of the density ratios
observed for collisionless shocks in the heliosphere (upper panel
in Figure~9).

\section{Conclusions}
\begin{enumerate}
\item A brightness discontinuity, interpreted as a shock wave, was
demonstrated to be registered in a limited region in front of each
of the ten selected CMEs (propagating in the plane of the sky)
along their travel directions. In various events, the shock wave
velocity was $V\approx$ 800-2500 km s$^{-1}$.
\item Comparing the dependence of the Alfv{\'e}n Mach number $M_A$ on the shock
wave strength $\rho_2/\rho_1$, measured at $R > 10R_\odot$ from
the center of the Sun, with ideal MHD calculations suggests that
the effective adiabatic index $\gamma$, characterizing the
processes inside the front is largely between 2 and 5/3. This
corresponds to the effective number of degrees of freedom of
motion 2 to 3. A similar dependence $M_A(\rho_2/\rho_1)$ was
obtained for Earth's bow and interplanetary collisionless shock
waves. All these substantiate the assumption that the
discontinuities in question in front of the CMEs at $R >
10R_\odot$ are collisionless shock waves.
\end{enumerate}

%

%

%
 \appendix

\section{Deriving the dependence of the Alfv{\'e}n Mach number on
the density ratio at the flat shock using the ideal MHD
approximation}

\subsection{Notations and initial equations}
Let us consider a flat shock front. Index 1 will denote the
undisturbed plasma region ahead of the front, index 2 the region
behind the front. A front-associated coordinate system is used,
i.e. the plasma velocity ahead of the front is the same as the
front velocity. The front surface is in the $yz$ plane, and the
$x$ axis is perpendicular to the surface. The coordinate system is
chosen such that the magnetic field vectors and velocities lie in
the $xy$ plane. The magnetic field vector $B_1$ is at an angle,
$\theta_{Bn}$, to the $x$ axis (i.e. to the normal to the front).
Thus, the case of $\theta_{Bn} = 90^\circ$ corresponds to the
perpendicular shock when the magnetic field vector is in the plane
of the front.

According to \opencite{Priest1982}, the following conditions are
satisfied at the shock discontinuity (in the CGS system):
\begin{equation}
\rho_2v_{2x}=\rho_1v_{1x}
\end{equation}
\begin{equation}
p_2+B_2^2/8\pi-B_{2x}^2/4\pi+\rho_2v_{2x}^2=p_1+B_1^2/8\pi-B_{1x}^2/4\pi+\rho_1v_{1x}^2
\end{equation}
\begin{equation}
\rho_2v_{2x}v_{2y}-B_{2x}B_{2y}/4\pi=\rho_1v_{1x}v_{1y}-B_{1x}B_{1y}/4\pi
\end{equation}
\begin{eqnarray}
(p_2+B_2^2/8\pi)v_{2x}-B_{2x}({\bf B_2\cdot
v_2})/4\pi+(\rho_2e_2+\rho_2v_2^2/2+B_2^2/8\pi)v_{2x}=\nonumber \\
(p_1+B_1^2/8\pi)v_{1x}-B_{1x}({\bf B_1\cdot
v_1})/4\pi+(\rho_1e_1+\rho_1v_1^2/2+B_1^2/8\pi)v_{1x}
\end{eqnarray}
\begin{equation}
B_{2x}=B_{1x}
\end{equation}
\begin{equation}
v_{2x}B_{2y}-v_{2y}B_{2x}=v_{1x}B_{1y}-v_{1y}B_{1x}
\end{equation}
here $v$ is velocity, $B$ is magnetic field, $p$ is pressure,
$\rho$ is density, $e$ is internal energy defined by
$e=\displaystyle\frac{p}{(\gamma-1)\rho}$ for the ideal polytropic
gas, where $\gamma$ is the adiabatic index.

Equation (1) is the mass conservation condition; equations (2) and
(3) are the conservation conditions respectively of the $x$ and
$y$ components of the momentum; equation (4) is the energy
conservation condition. Equation (5) (following from div ${\bf B}
= 0$) is the conservation condition of the normal component $B_x$
of the magnetic field. Equation (6) is the continuity condition of
the [${\bf v}\times {\bf B}$] value following from the electric
field tangential component continuity and magnetic field
freezing-in condition.

Equations are written here in the general form, but the velocity
and magnetic field $z$-components are zero throughout the selected
coordinate system. At the shock discontinuity, the velocity
changes -- only the $x$ component of velocity
is present in region 1, whereas the $y$ component is
present in region 2 as well. The magnetic field, namely its $y$
component, also varies in the front, whereas the $x$ component
remains unchanged.

\subsection{Derivation and solution of main equations}

Let us introduce the notation for the ratio of tangential (about a
wave front) magnetic field components:
\begin{displaymath}
h=B_{2y}/B_{1y},
\end{displaymath}
and the notation:
\begin{displaymath}
\Gamma = \gamma/(\gamma-1).
\end{displaymath}
We will additionally use the following expressions:
$v_A=B/\sqrt{4\pi\rho}$ for the Alfv{\'e}n velocity, $M_A =
v_{1x}/v_{A_1}$ for the Alfv{\'e}n Mach number, and $\beta =
\displaystyle\frac{p}{B^2/8\pi}$ for the gas to magnetic pressure
ratio.

Equation (3), with due account for (1) and (5), may be used to
express the tangential velocity component appearing behind the
front:
\begin{displaymath}
v_{2y}=\frac{B_1}{4\pi\rho_1v_{1x}}(h-1)\cos^2\theta_{Bn}\tan\theta_{Bn}
\end{displaymath}

Substituting the $v_{2y}$ expression into (6) and taking (1) into
account, we get:
\begin{equation}
\frac{v_{2x}}{v_{1x}} = \frac{\rho_1}{\rho_2} = \frac{1}{h}\,
(1+\frac{h-1}{M^2_A}\, \cos^2\theta_{Bn} )
\end{equation}

The expression (7) relates the density ratio $\rho_2/\rho_1$ and
the Alfv{\'e}n Mach number $M_A$. However, this expression also
includes the ratio between magnetic field tangential components,
$h$; thus we must construct and solve an equation system to
express $M_A$ in terms of $h$.

Dividing both sides of Equation (2) by $\rho_1v_{1x}^2$ with due
account for (5) and expression for $v_{2x}/v_{1x}$ yields:
\begin{equation}
\frac{1}{h}\, (1+\frac{h-1}{M^2_A}\, \cos^2\theta_{Bn} )\, (1 +
\alpha_2) = \frac{\beta_1}{2M_A^2} +
\frac{(1-h^2)\sin^2\theta_{Bn}}{2M_A^2}+1
\end{equation}
We introduce the notation $\alpha_2 =
\displaystyle\frac{p_2}{\rho_2v_{2x}^2}$ here.

Dividing both sides of Equation (4) by
$\displaystyle\frac{\rho_1v_{1x}^3}{2}$ and substituting
expressions for $v_{2y}$ and $v_{2x}/v_{1x}$, we obtain a second
equation:
\begin{eqnarray}
(1 + 2\Gamma \alpha_2)\, \frac{1}{h^2}\, (1+\frac{h-1}{M^2_A}\,
\cos^2\theta_{Bn})^2 + \frac{2(h-1)\sin^2\theta_{Bn}}{M_A^2}
\nonumber\\ + \frac{(h-1)^2\sin^22\theta_{Bn}}{4M_A^4}=
\frac{\Gamma \beta_1}{M_A^2}+1
\end{eqnarray}

Thus there are two equations to express $M_A$ in terms of $h$.
Expressing $\alpha_2$ from the first equation and substituting it
into the second one produces, after some simplification:
\begin{eqnarray}
\frac{h-1}{hM_A^2}\,
\Bigl(\bigl(M_A^2-\cos^2\theta_{Bn}\bigr)\bigl(M_A^2(1+h-2\Gamma)+h\beta_1\Gamma-
(h-1)(2\Gamma-1)\cos^2\theta_{Bn}\bigr)\nonumber \\
+h\sin^2\theta_{Bn}\bigl(M_A^2(h(\Gamma-2)+\Gamma)+(h-1)(h(\Gamma-1)+\Gamma)\cos^2\theta_{Bn}\bigr)\Bigr)=0\nonumber
\\
\end{eqnarray}
This equation has two solutions for $M_A^2$:
\begin{eqnarray}
M_A^2=\frac{1}{2(1+h-2\Gamma)}\,
\biggl(-h\beta_1\Gamma+2(1+(h-2)\Gamma)\cos^2\theta_{Bn}
\hspace{20mm} \nonumber\\
 -h(h(\Gamma-2)+\Gamma)\sin^2\theta_{Bn}
\nonumber\\ \pm \sqrt{h^2\bigl(\, 4(\Gamma-1)^2\cos^4\theta_{Bn}
+(\beta_1\Gamma+(h(\Gamma-2)+\Gamma)\sin^2\theta_{Bn})^2}
\hspace{1mm} \nonumber \\ \overline{\vphantom{\Bigl(}
-4(\Gamma-1)\cos^2\theta_{Bn}(\beta_1\Gamma+(1+h^2-(1+h)\Gamma)\sin^2\theta_{Bn}
)\, \bigr)}\; \biggr)
\end{eqnarray}

The solution with `$+$' before the root gives the trivial value
$M_A^2 = 0$ at $\theta_{Bn} = 90^\circ$ and physically infeasible
values $M_A^2 < 0$ at $\theta_{Bn} \neq 90^\circ$. Of interest is
therefore the second solution with `$-$' before the root.

\subsection{Extreme cases}

Let us consider some extreme cases of the solution. We may set
$\beta_1 = 0$ for plasma totally controlled by the magnetic field,
and $\gamma = 5/3$ and correspondingly $\Gamma = 2.5$ for gas with
three degrees of freedom.

In the perpendicular shock case ($\theta_{Bn} = 90^\circ$), the
expression for $M_A^2$ is simplified and takes the form:
\begin{displaymath}
M_A^2=\frac{h(2.5+0.5h)}{4-h}
\end{displaymath}
If $M_A^2 = 1$, then $h = 1$, and if $M_A^2 \rightarrow \infty$,
then $h\rightarrow 4$. As is evident from (8), the density ratio
in the shock front in this case simply equals the magnetic field
tangential component ratio, i.e. $\rho_2/\rho_1 = h$.

In the parallel shock case ($\theta_{Bn} = 0^\circ$), $M_A^2 = 1$
for any $h$ because it is an extreme case of shock wave unaffected
by the magnetic field.

In the general case of $\beta_1\neq0$, $M_A^2 > 1$ for $h = 1$
because the characteristic velocity of disturbance propagation in
the plasma will be the fast magnetosonic wave velocity rather than
the Alfv{\'e}n velocity $v_A$ \cite{Priest1982}
\begin{displaymath}
v_{MS}= \sqrt{\frac{1}{2}\,
(c_s^2+v_A^2)+\frac{1}{2}\sqrt{c_s^4+v_A^4-2c_s^2v_A^2\cos2\theta_{Bn}}}
\end{displaymath}
where $c_s$ is the sound velocity. In this case it is more
correctly to try to find the dependence of the magnetosonic Mach
number $M_{MS} = v/v_{MS}$ on density and magnetic field ratios.

%
 \begin{acks}
The work was supported by the Program No. 16 part 3 of the
Presidium of the Russian Academy of Sciences, Program of State
Support for Leading Scientific Schools NS-2258.2008.2, and the
Russian Foundation for Basic Research (Project No. 09-02-00165a).
The SOHO/LASCO data used here are produced by a consortium of the
Naval Research Laboratory (USA), Max-Planck-Institut fuer
Aeronomie (Germany), Laboratoire d'Astronomie (France), and the
University of Birmingham (UK). SOHO is a project of international
cooperation between ESA and NASA.
 \end{acks}

%
%
%

\end{article}
\end{document}